\documentclass[a4paper,10pt]{article}
\usepackage{jheppub}
\usepackage{hyperref}

\usepackage{graphicx}
\usepackage{dcolumn}
\usepackage{bm}

\usepackage{enumitem}
\usepackage[dvipsnames]{xcolor}
\definecolor{codepurple}{rgb}{0.58,0,0.82}
\definecolor{Red}{rgb}{0,0,0} 

\usepackage[bottom]{footmisc}

\usepackage{float}
\usepackage{csquotes}
\usepackage{tensor}
\usepackage{mathtools}
\usepackage{physics}
\usepackage{cancel}

\colorlet{linkequation}{codepurple}
\newcommand*{\SavedEqref}{}
\let\SavedEqref\eqref
\renewcommand*{\eqref}[1]{%
  \begingroup
    \hypersetup{
      linkcolor=linkequation,
      linkbordercolor=linkequation,
    }%
    \SavedEqref{#1}%
  \endgroup
}

\definecolor{codepurple}{rgb}{0.58,0,0.82}
\usepackage{csquotes}
\usepackage{tensor}
\usepackage{mathtools}
\usepackage{physics}
\usepackage{cancel}
\usepackage{makecell}
\usepackage{etoolbox}
\usepackage{comment}
\usepackage{footnpag}

\usepackage[
]{footnotehyper}
\usepackage{chngcntr}

\counterwithout{footnote}{section}

\makeatletter
\renewcommand*{\@makefnmark}{\textsuperscript{\arabic{footnote}}}
\makeatother

\usepackage{slashed}
\usepackage{braket}
\usepackage{amssymb}

\begin{document}

\title{Gravitational and electromagnetic Cherenkov radiation constraints in modified dispersion relations}

\author{Mikel Artola, José A. R. Cembranos, Prado Martín-Moruno}
\affiliation{\vspace{0.3cm} Departamento de Física Teórica and Instituto de Física de Partículas y del Cosmos (IPARCOS-UCM), 
Facultad de Ciencias Físicas, 
Universidad Complutense de Madrid, E-28040 Madrid, Spain}

\emailAdd{mikart01@ucm.es}
\emailAdd{cembra@ucm.es}
\emailAdd{pradomm@ucm.es}

\date{\today}

\abstract{
Motivated by different approaches to quantum gravity, one could consider that Lorentz invariance is not an exact symmetry of nature at all energy scales. Following this spirit, modified dispersion relations have been used to encapsulate quantum gravity phenomenology. In the present work, we propose a class of Lorentz invariance violating phenomenological dispersion relations, which could be different for each particle species, to study the generalized vacuum Cherenkov radiation process. We identify the kinematic regions where the process is allowed and then compute the energy loss rate due to the emission of vacuum electromagnetic and gravitational Cherenkov radiation. Furthermore, we estimate constraints for the Lorentz invariance breaking parameters of protons and gravitons taking into account the existence (or absence) of vacuum gravitational Cherenkov radiation using ultra high energy cosmic ray detections.
}

\preprint{IPARCOS-UCM-24-044}

\maketitle

\section{Introduction}\label{sec_I}
%
%
Electromagnetic Cherenkov radiation occurs when an electrically charged particle traveling through an optical medium has a velocity $v$ larger than the phase velocity of light in that medium $c_\gamma$. %
This phenomenon was first detected by Cherenkov in 1934 \cite{Cherenkov_1934}, and a theoretical explanation was given by Frank and Tamm three years later \cite{Tamm_1937}. %
Cherenkov radiation (CR) can be understood microscopically in the following way \cite{Schreck_2018}. When the particle travels at a velocity $v \leq c_\gamma$, the polarized atoms close to the trajectory emit out of phase radiation, leading to destructive interference. %
However, when the particle has a velocity $v > c_\gamma$, the wave trains emitted by atoms and molecules are in phase, and the constructive interference taking place in a Mach cone, known as the Cherenkov cone, results in the coherent radiation observed in the direction perpendicular to the cone. %
Likewise, this phenomenon can be interpreted both qualitatively and quantitatively as an emission process $a \to a + \gamma$ where a charged particle $a$ emits a photon $\gamma$: the particle outruns the electromagnetic field, causing the emission of radiation because of the accumulation of wavefronts propagating from the particle \cite{Schreck_2018, Moore_2001}. %
The rate of energy loss in an optical medium with refractive index $n$ is described classically by the Frank-Tamm formula \cite{Tamm_1937}, and receives small corrections when considering quantum effects \cite{cox1944momentum}. %
Moreover, this formula can also be obtained from the quantum field theory formalism \cite{Moore_2001}. %
Anyway, for the process to be allowed the velocity of the particle must be greater than the phase velocity of light and, therefore, Cherenkov emission is not possible in vacuum in the framework of special relativity (SR) since $c_\gamma = c$.\par
%
%
Nevertheless, over the last decades there has been a growing interest in studying theoretical frameworks which suggest that Lorentz invariance (LI) may not be an exact symmetry of nature at all energy scales \cite{Mattingly_2005}. %
Such theories appear mainly in the context of quantum gravity (QG) 
\cite{Addazi_2022,batista2023white} where, if LI is violated, the Planck energy $E_\text{Pl} \approx 10^{19}\, \text{GeV}$ is expected to be the scale where this symmetry is strongly violated. %
Although there is a large energy gap between the highest energy particles detected, those are ultra high energy cosmic rays (UHECRs) $\approx 10^{11}\, \text{GeV}$, and with the Planck scale, there should be an interpolation of LI violation (LIV) to the low energy regime, where these particles could be sensitive to small departures from LI. %
Focusing on astrophysical phenomena, the most straightforward way to implement LIV is to consider modified dispersion relations (MDRs) for particles maintaining the 4-momentum conservation laws. %
In this scenario, kinematics of a wide variety of processes may be affected in what are called threshold effects depending on the MDRs considered. These effects could entail the shift of the energy thresholds or add new thresholds to processes already predicted in the LI scenario, or even allow completely new reactions.\par
%
%
Vacuum CR may be allowed in a LI violating frame, and it has been exhaustively studied in the electromagnetic sector \cite{Coleman_1997,Coleman_1999,Jacobson_2003,Jacobson_2006,Gagnon_2004} to impose restrictive constraints to LI violating parameters (LIVPs) using high energy astrophysical particles. %
This idea has also been extended to study the vacuum gravitational CR \cite{Moore_2001,Elliott_2005,Kimura_2012,Kiyota_2015,Kosteleck_2015}, where an arbitrary particle can lose energy due to the emission of gravitons rather than photons. %
With the recent observations of gravitational waves (GWs) \cite{Yunes_2016, Abbott_2017}, combined constraints for LIVPs were obtained in the gravitational sector. %
These studies, however, only modify the dispersion relation in the gravitational sector or the matter sector, but combined effects have not been investigated yet with nontrivial MDRs. %
The main aim of the present paper is to consider MDRs both in the gravitational and the matter sector simultaneously to obtain combined restrictions in the LIVPs taking into account the existence or absence of vacuum gravitational CR.\par
%
%
The remainder of this work is structured as follows. %
In Sec. \ref{sec_II} we propose a phenomenological MDR and study the kinematics of the generalized Cherenkov radiation (GCR) process $a \to a + b$. %
In Sec. \ref{sec_III} we apply the results obtained in Sec. \ref{sec_II} to compute the rate of energy loss of the particle (Sec. \ref{sec_III_a}) for vacuum electromagnetic and gravitational CR (Secs. \ref{sec_III_b} and \ref{sec_III_c}). %
Sec. \ref{sec_IV} is devoted to impose constraints in LIVPs using UHECRs, and, finally, in Sec. \ref{sec_V} we discuss the main results obtained and its limitations. %
Along this work we use natural units, i.e., we set $\hbar = c = 1$.

\section{Kinematics of generalised Cherenkov radiation}\label{sec_II}
%
%
\subsection{General aspects of the modified dispersion relation}
%
%
Nowadays it is not yet clear how LI might be broken, if it is the case. However, there are hints about the kind of phenomenology it could produce. An interesting example
is the possibility of new particle decays due to LIV. %
In order to study the kinematics of these processes, it has previously been proposed \cite{Coleman_1999} that each particle $a$ has, besides its own mass $m_a$, its own maximum velocity $c_a$, which is asymptotically achievable if the particle has a nonvanishing mass. %
The dispersion relation in this scenario is given by
\begin{equation}\label{eq:mdr_coleman}
    E_a^2 = m_a^2 c_a^4 + \boldsymbol{p}_a^2 c_a^2,
\end{equation}
where $E_a$ and $\boldsymbol{p}_a$ are the energy and 3-momentum of the particle, respectively. %
On the other hand, the Cherenkov effect is usually studied in terms of the refractive index $n$ of an optical medium, which modifies the maximum attainable velocity of light in vacuum $c$ as $c_\gamma = c n^{-1}$. %
Hereafter we will set $c = 1$. %
Following this idea, we consider that it is convenient to parametrize the deviation of the maximum attainable velocity $c_a$ for each particle species in terms of a particular refractive index $n_a$, so that $c_a = n_a^{-1}$ and Eq. \eqref{eq:mdr_coleman} can be written as
\begin{equation}\label{eq:mdr_original}
    E_a^2 = m_a^2 n_a^{-4} + \boldsymbol{p}_a^2 n_a^{-2}.
\end{equation}
The existence of this refractive index $n_a$ for particles of the Standard Model can be interpreted in the context of quantum gravity, for instance, as the spacetime foam acting as a medium \cite{Ellis_2008}. In a similar manner, for GWs it may act as a diagravitational medium \cite{Cembranos_2019}.\par
%
%
In this work we propose a phenomenological dispersion relation where the refractive index $n_a$ of a particle $a$ can depend on its energy as a power law
\begin{equation}\label{eq:index_of_refraction}
    n_a = 1 + \mathbb{A}^{(\alpha)} E_{\boldsymbol{p}_a}^\alpha,
\end{equation}
where there is no summation in $\alpha$ as we take into account only the dominant term in the deviation from unity. %
Both the {\color{Red} constant} coefficient $\mathbb{A}^{(\alpha)}$, with units $[\mathbb{A}^{(\alpha)}] = [E^{-\alpha}]$, and the exponent $\alpha$ are allowed to be different for each particle species. %
{\color{Red} To illustrate the utility of Eq. \eqref{eq:index_of_refraction}, we point out that when $\alpha = 0$, some MDRs predicted by different theoretical frameworks can be described, e.g. the setting characterized by  isotropic coefficients in the photon sector found in \cite{Klinkhamer_2008} or in the fermion sector studied in \cite{Schreck_2017}.}
From now on we shall denote $\mathbb{A} \equiv \mathbb{A}^{(\alpha)}$ except when discussing different values of $\alpha$. %
As LI is almost an exact symmetry we will assume that $|\mathbb{A} E_{\boldsymbol{p}_a}^\alpha| \ll 1$. %
Replacing the refractive index \eqref{eq:index_of_refraction} in the MDR \eqref{eq:mdr_original} and considering terms up to first order in $|\mathbb{A} E_{\boldsymbol{p}_a}^\alpha|$ we get
\begin{equation}\label{eq:MDR_order_1}
    E_{\boldsymbol{p}_a}^2 =
    m_a^2 + \boldsymbol{p}_a^2 - 2\mathbb{A} E_{\boldsymbol{p}_a}^\alpha (2m_a^2 + \boldsymbol{p}_a^2).
\end{equation}
Even though $m_a$ and $|\mathbb{A} E_{\boldsymbol{p}_a}^\alpha|$ are independent scales, we will see that it is safe to expand only in the LI violating term for high energy particles, hence no multiscale problems arise from this analysis. %
Note that Eq. \eqref{eq:MDR_order_1} is invariant under translations and rotations but not under Lorentz boosts due to the dependence of $n_a$ on the energy of the particle. %
Thus, this MDR only holds in a particular preferred frame, which is usually chosen to be at rest with respect to the CMB. %
The relative velocity between Earth and the CMB is of order $\mathcal{O}(10^{-3})$ \cite{Planck_2018} and its effects can be neglected when imposing constraints on the LIVPs.\par
%
%
It should be noted that for the MDR given by Eq. \eqref{eq:mdr_original} the group velocity and the phase velocity are different in general. %
The maximum attainable velocity of the particle $c_a = n_a^{-1} = 1 - \mathbb{A}p^\alpha + \mathcal{O}(\mathbb{A}^2)$ is referred to the phase velocity, which differs from the group velocity defined as $v_a(E) = \partial E/\partial p$. %
Indeed, for high energy particles with $p \gg m$ one has, using Eq. \eqref{eq:MDR_order_1} and neglecting the mass in the last term,
\begin{equation}\label{eq:velocity}
    v_a = 1 - \frac{m^2}{2p^2} - (\alpha + 1)\mathbb{A}p^\alpha + \mathcal{O}(\mathbb{A}^2),
\end{equation}
where $p$ is the module of the 3-momentum of the particle. %
We then see that the maximum achievable group velocity, $v_a(m = 0) = 1 - (\alpha + 1)\mathbb{A}\,p^\alpha$, is different from that given by $c_a$ when $\alpha \neq 0$. %
As we will see, for the processes we want to study, the kinematics is governed by the relation between the maximum attainable phase velocities $c_a$ of the particles, or, equivalently, the relation between the refractive indexes $n_a$.\par
\begin{figure}[t]
    \centering
    \includegraphics[width=0.4\textwidth]{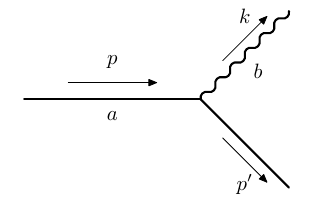}
    \vspace{-0.25cm}
    \caption{Tree-level GCR process where a particle $a$ (continuous line) emits a massless particle $b$ (wiggly line).}
    \label{fig:generalized_cherenkov}
\end{figure}

%
%
The MDR \eqref{eq:MDR_order_1} could be used to study the shift in the threshold energy or new thresholds of existing reactions. %
If the mass of the particle is high enough compared to the LI violating term, then the usual dispersion relation holds and kinematics remain, essentially, unmodified. %
However, if the LI breaking term is comparable to the mass of the particle, significant deviations are expected in the kinematics of a wide variety of reactions. %
As the thresholds of the processes are determined by the mass of the particles, modifications to those thresholds are appreciable when the third term in the MDR \eqref{eq:MDR_order_1} is of the same order as the mass $m_a$. For high energy particles $p \gg m$ the energy $p_\text{dev}$ at which the deviation becomes important can be estimated by
\begin{equation}\label{eq:deviation_threshold}
    p_\text{dev} \sim \left| \frac{m^2}{2\mathbb{A}} \right|^{1/(\alpha + 2)}.
\end{equation}
In this work we are interested in studying the existence of entirely new reactions that are forbidden in SR; in particular we will consider the GCR process (see Fig. \ref{fig:generalized_cherenkov}), i.e., the two body decay process $a \to a + b$ where a particle $a$ emits a massless particle $b$%
\footnote{Here we mention that forbidden processes in the LI scenario cannot be studied in the scope of doubly special relativity (DSR). %
This is because the deformed 4-momentum composition produces cancellations with the LI violating terms in such a way that this phenomenology is not possible \cite{Addazi_2022}.}.%

%
%
%
%
\subsection{Derivation of the Cherenkov angle}
One of the key ingredients when studying CR is the angle of emission $\theta_c$ between the 3-momentum $\boldsymbol{p}$ of the initial particle $a$ and the 3-momentum $\boldsymbol{k}$ of the emitted particle $b$. %
If we define $E_{\boldsymbol{p}} := E_a(\boldsymbol{p})$, $E_{\boldsymbol{p^\prime}} := E_a(\boldsymbol{p}^\prime)$ and $E_{\boldsymbol{k}} := E_b(\boldsymbol{k})$, where $\boldsymbol{p}^\prime$ is the 3-momentum of the particle $a$ after the emission (see Fig. \ref{fig:generalized_cherenkov}), then the conservation of energy and 3-momentum, which are written as
\begin{equation}\label{eq:conservation_4-momentum}
    E_a(\boldsymbol{p}) = E_b(\boldsymbol{k}) + E_a(\boldsymbol{p}^\prime)
    ,\hspace{15pt}
    \boldsymbol{p} = \boldsymbol{k} + \boldsymbol{p}^\prime,
\end{equation}
allow us to obtain $\theta_c$ in terms of the module of the $3$-momenta $\boldsymbol{p}$ and $\boldsymbol{k}$. %
Note that Eq. \eqref{eq:conservation_4-momentum} assumes the usual 4-momentum composition; hence, DSR theories are excluded from this analysis.\par
%
%
Let us now use the MDR given by Eq. \eqref{eq:mdr_original} with refractive index \eqref{eq:index_of_refraction} to obtain the Cherenkov angle $\theta_c$ or, equivalently, $\cos{\theta_c}$. %
Let $m$ be the mass of particle $a$ (we remind the reader that $b$ is massless) and denote $n_{\boldsymbol{p}} := n_a(\boldsymbol{p})$, $n_{\boldsymbol{p} - \boldsymbol{k}} := n_a(\boldsymbol{p} - \boldsymbol{k})$ and $n_{\boldsymbol{k}} := n_b(\boldsymbol{k})$, where we already applied the conservation of 3-momentum. %
From Eq. \eqref{eq:mdr_original}, we see that the energies of the particles in the GCR process are given by
\begin{align}
        E_{\boldsymbol{p}}^2 &=
        m^2 n_{\boldsymbol{p}}^{-4} + p^2 n_{\boldsymbol{p}}^{-2}, \notag \\
        E_{\boldsymbol{p} - \boldsymbol{k}}^2 &=
        m^2 n_{\boldsymbol{p} - \boldsymbol{k}}^{-4} + (p^2 + k^2 - 2pk \cos{\theta_c}) n_{\boldsymbol{p} - \boldsymbol{k}}^{-2}, \label{eq:energies_GCR} \\
        E_{\boldsymbol{k}}^2 &=
        k^2 n_{\boldsymbol{k}}^{-2}, \notag
\end{align}
where we have used $(\boldsymbol{p} - \boldsymbol{k})^2 = p^2 + k^2 - 2pk \cos{\theta_c}$ and denoted the module of the 3-momentum of the particles $a$ and $b$ as $p$ and $k$, respectively. %
On the other hand, combining the conservation of energy and 3-momentum we have $E_{\boldsymbol{p} - \boldsymbol{k}}^2 = [E_{\boldsymbol{p}} - E_{\boldsymbol{k}}]^2$. Substituting here the energies given by Eq. \eqref{eq:energies_GCR}, we get
\begin{eqnarray}\label{eq:cos_1}
    2pk n_{\boldsymbol{p} - \boldsymbol{k}}^{-2} \cos{\theta_c} &=&
    p^2 (n_{\boldsymbol{p} - \boldsymbol{k}}^{-2} - n_{\boldsymbol{p}}^{-2}) + m^2 (n_{\boldsymbol{p} - \boldsymbol{k}}^{-4} - n_{\boldsymbol{p}}^{-4}) + k^2 (n_{\boldsymbol{p} - \boldsymbol{k}}^{-2} - n_{\boldsymbol{k}}^{-2}) \nonumber\\
    &+& 2 n_{\boldsymbol{k}}^{-1} k \left[ n_{\boldsymbol{p}}^{-2} p^2
    + n_{\boldsymbol{p}}^{-4} m^2 \right]^{1/2}.
\end{eqnarray}
Note that in the LI scenario, which corresponds to $n_a = n_b = 1$, the rhs of Eq. \eqref{eq:cos_1} is always greater than the lhs unless $\cos{\theta_c} > 1$; hence, the process is forbidden. The limit case where the particle $a$ has no mass is of no interest when $n_a = n_b = 1$, as the emission rate, which we compute in Sec. \ref{sec_III}, vanishes for $\cos{\theta_c} = 1$. \par
Now let $\mathbb{A}$ and $\alpha$ be the LIVP and exponent of the refractive index of the particle $a$, and $\mathbb{B}$ and $\beta$ those of the particle $b$, such that
\begin{equation}
    n_a(\boldsymbol{p}) =
    1 + \mathbb{A} E_a^\alpha(\boldsymbol{p})
    ,\hspace{15pt}
    n_b(\boldsymbol{k}) =
    1 + \mathbb{B} E_b^\beta(\boldsymbol{k}).
\end{equation}
We remark that for the massless particle we are ignoring a small mass term $m_b$, and terms that induce birefringence or attenuation, which could be present in alternative theories of gravity for the graviton \cite{Cembranos_2019} or in the Standard Model extension for gauge bosons. %
Assuming that $|\mathbb{A} E_{\boldsymbol{p}}^\alpha|, |\mathbb{A} E_{\boldsymbol{p}-\boldsymbol{k}}^\alpha|, |\mathbb{B} E_{\boldsymbol{k}}^\alpha| \ll 1$ we can replace these refractive indexes in Eq. \eqref{eq:cos_1} and expand up to first order to obtain
\begin{eqnarray}\label{eq:cos_2}
    2pk\cos{\theta} &=&
    4pk \mathbb{A} E_{\boldsymbol{p} - \boldsymbol{k}}^\alpha \cos{\theta} -
    2\mathbb{A}(p^2 + 2m^2) [E_{\boldsymbol{p} - \boldsymbol{k}}^\alpha - E_{\boldsymbol{p}}^\alpha] \nonumber\\ 
    &+&
    2k (p^2 + m^2)^{1/2} \left[ 1 - \mathbb{B} E_{\boldsymbol{k}}^\beta \right] \left[ 1 - 2\mathbb{A} \frac{(p^2 + 2m^2) E_{\boldsymbol{p}}^\alpha }{p^2 + m^2} \right]^{1/2} \nonumber\\ 
    &-&
    2k^2 [\mathbb{A} E_{\boldsymbol{p} - \boldsymbol{k}}^\alpha - \mathbb{B} E_{\boldsymbol{k}}^\beta].
\end{eqnarray}
This is an implicit Eq. in $\cos{\theta}$ as the energy $E_{\boldsymbol{p} - \boldsymbol{k}}$ depends on $(\boldsymbol{p} - \boldsymbol{k})^2$, which in turn depends on $\cos{\theta}$.\par
%
%
In Sec. \ref{sec_IV} we will consider UHECR observations to constrain the LI breaking parameters $\mathbb{A}$ and $\mathbb{B}$. %
For these particles $p \gg m$. %
In addition, it can be checked \emph{a posteriori} that the threshold is greatly modified, that is, the momentum is much larger than the right-hand side of Eq. \eqref{eq:deviation_threshold}. %
We can then consider terms up to $\mathcal{O}(m^2/p^2)$ and neglect first order products of the LI violating terms with the quotient $m^2/p^2$. %
Applying these additional approximations in Eq. \eqref{eq:cos_2} we get
\begin{equation}\label{eq:cos_final}
    \cos{\theta_c} := 1 - \Theta_c
\end{equation}
where
\begin{align}\label{eq:Theta_1}
    \Theta_c :=
    \frac{p - k}{pk} \left\{
    (p - k) \mathbb{A} E_{\boldsymbol{p} - \boldsymbol{k}}^\alpha +
    \mathbb{B} k E_{\boldsymbol{k}}^\beta - \mathbb{A} p E_{\boldsymbol{p}}^\alpha \right\} - \frac{m^2}{2p^2}.
\end{align}
As $\Theta_c$ is already of first order in the LI violating terms and $m^2/p^2$, then the three-vectors $\boldsymbol{p}$ and $\boldsymbol{k}$ involved in the energies must be considered collinear and no second order corrections are taken into account. %
In other words, $\Theta_c$ does not depend on $\cos{\theta_c}$ and, consequently, Eq. \eqref{eq:cos_final} completely characterizes the Cherenkov angle $\theta_c$ in terms of the initial 3-momentum of the particle $a$ and the 3-momentum carried off by the particle $b$.\par
%
%
\subsection{Threshold condition}\label{sec:II_b}
%
%
Attending to Eq. \eqref{eq:cos_final}, the Cherenkov process $a \to a + b$ is allowed as long as $\Theta_c> 0$ such that $\cos{\theta_c} < 1$. Hence the threshold is determined by the equation
\begin{equation}\label{eq:threshold}
    \Theta_c = 0.
\end{equation}
The study of the threshold condition is easier to carry out if the energies in Eq. \eqref{eq:Theta_1} are substituted by the momentum, which is correct up to first order. Thus, we obtain
\begin{eqnarray}
    \Theta_c &=&
    \frac{p - k}{p}
    f(k) - \frac{m^2}{2p^2}, \label{eq:Theta_3} \\
    f(k) &:=&
    \mathbb{B} k^\beta - \frac{\mathbb{A}}{k} \left\{ p^{\alpha + 1} - (p - k)^{\alpha + 1} \right\}. \label{eq:f(k)}
\end{eqnarray}
In what follows we will consider $k \leq p$, which is satisfied particularly when studying the threshold condition. %
Indeed, if $k > p$, then the massless particle would be anti-parallel to the incoming particle, but this is forbidden by the threshold theorem \cite{Mattingly_2003}: if $E_{\boldsymbol{p}}$ is a strictly monotonically increasing function of $p$ for $p > 0$ for all particles, then all thresholds for processes with two particle final states occur when the final momentum are parallel. %
This theorem has to be satisfied in our case since, in the first place, as we have previously discussed, $E_{\boldsymbol{p}}$ is a rotational-invariant function of $\boldsymbol{p}$, so it is only a function of the modulus of the 3-momentum $p$; in the second place, it is a strictly monotonically increasing function of $p$, as the LI violating terms are much smaller than the momentum of the particle.\par 
%
%
Before focusing on the threshold condition, it is possible to establish whether the process is allowed or forbidden attending to the signs of the parameters $\mathbb{A}$ and $\mathbb{B}$. %
Two cases can be studied without loss of generality.
\begin{enumerate}
    \item \underline{$\mathbb{A} > 0$ and $\mathbb{B} < 0$.} %
    It is easy to see from Eq. \eqref{eq:Theta_3} that $\Theta_c < 0$ for all $k < p$, as the energy of the incoming particle is always greater than the energy of the particles in the final state. %
    Hence, Cherenkov emission is forbidden. %
    In this scenario $n_a > n_b$ and, therefore, $c_a<c_b$. So, the intuitive condition that the particle $a$ must have a greater phase velocity than the particle $b$ in order for the process to be allowed is never fulfilled.
    \item \underline{$\mathbb{A} < 0$ and $\mathbb{B} > 0$.} In this case we see that all the terms in Eq. \eqref{eq:Theta_3} are positive except the correction $m^2/(2p^2)$ due to the mass of the particle. %
    For particles with energies many orders of magnitude greater than the one established by the rhs of Eq. \eqref{eq:deviation_threshold}, this term might be neglected and, thus, Cherenkov emission is allowed for $k \lesssim p$. %
    Now we have that $n_a < n_b$ ($c_a>c_b$), which fits with the idea that the process is possible since the particle $a$ has a greater phase velocity than the particle $b$.
\end{enumerate}
%
%
For the remaining two cases, the sign of $\Theta_c$ depends on the signs of $\mathbb{A}$ and $\mathbb{B}$ but also on the momenta $p$ and $k$. %
It is then necessary to obtain the threshold momentum of the emitted particle $b$, namely $k_\text{th}$, in order to determine the values of $k$ for which the process is allowed. %
We remark that the following results of this Sec. are valid for $\alpha \geq 0$ and $\beta \geq 0$%
\footnote{Recent theoretical results \cite{Freidel_2021} showed that QG phenomenology in the infrared could be modified, but this has not been yet studied in the context of particles of astrophysical origin, primarily concerned with high energy physics. We thus not consider $\alpha, \beta < 0$ in the rest of the work.}.\par
%
%
Before doing so, we can study the relation between $\mathbb{A}$ and $\mathbb{B}$ when the emitted particle $b$ has arbitrarily low momentum, $k \to 0$, or carries off all of the momentum of the particle $a$, $k \to p$. %
In the former case, it is easy to see from Eq. \eqref{eq:Theta_3} that Cherenkov emission is possible if
\begin{equation}\label{eq:limit_k_0}
    \left\{
    \begin{array}{rc}
         \displaystyle \mathbb{B} - (\alpha + 1) \mathbb{A} p^\alpha \geq \frac{m^2}{2p^2}, &\hspace{1em} \beta = 0; \\[5pt]
         \displaystyle \mathbb{A} \leq - \frac{m^2}{2(\alpha + 1) p^{\alpha + 2}}, &\hspace{1em} \beta > 0. 
    \end{array}
    \right.
\end{equation}
Note that the first case corresponds to a constant refractive index $n_b$. The process is not allowed when $k = p$ and $m \neq 0$, but for sufficiently high energy particles such that the mass may be neglected the process is allowed as long as 
\begin{equation}\label{eq:limit_k_p}
    \mathbb{B} p^\beta - \mathbb{A} p^\alpha \geq 0.
\end{equation}
The threshold condition is recovered setting an equal sign in Eqs. \eqref{eq:limit_k_0} and \eqref{eq:limit_k_p}.\par
%
%
It is also possible to prove in the massless limit that if the threshold condition $f(k_\text{th}) = 0$ has a solution, then $k_\text{th}$ is unique. %
Let us first focus on the case where $\mathbb{A} > 0$ and $\mathbb{B} > 0$. We start by calculating the derivative of $f(k)$, which is given by
\begin{align}
    f^\prime(k) &=
    \beta \mathbb{B} k^{\beta - 1} + \mathbb{A}k^{-2} p^{\alpha + 1} g(k), \label{eq:derivative_f} \\
    g(k) &= 1 - \left( 1 - \frac{k}{p} \right)^\alpha \left( 1 + \frac{\alpha k}{p} \right).
\end{align}
It is easy to check that, for the physical momentum $k \in [0,p]$, $g^\prime(k) > 0$ and $g(0)$, $g(p) > 0$; hence $g(k) > 0$. %
The remaining terms in Eq. \eqref{eq:derivative_f} are positive, so we also have that $f^\prime(k) > 0$; in other words, $f(k)$ is a monotonically increasing function. %
Therefore, if $f(0) < 0$ and $f(p) > 0$, the solution $k_\text{th}$ is unique. The condition $f(0) < 0$ is satisfied when Eq. \eqref{eq:limit_k_0} does not hold, and $f(p) > 0$ when Eq. \eqref{eq:limit_k_p} is fulfilled. %
The threshold momentum $k_\text{th}$ here establishes the minimum momentum that particle $b$ must have in order to have Cherenkov emission with momentum $k \in [k_\text{th}, p]$, and it is permitted as long as $k_\text{th} < p$. %
The limit case $k_\text{th} = p$ occurs setting an equal sign in Eq. \eqref{eq:limit_k_p}.\par
%
%
The proof is similar when $\mathbb{A} < 0$ and $\mathbb{B} < 0$. %
In this case one can check that $f(k)$ is a monotonically decreasing function, and so there exists a unique solution $k_\text{th}$ when $f(0) > 0$ and $f(p) < 0$. %
The first condition is met when Eq. \eqref{eq:limit_k_0} is satisfied, and the second one when Eq. \eqref{eq:limit_k_p} is not fulfilled. %
Hence, $k_\text{th}$ establishes the maximum momentum that particle $b$ can have up to where the Cherenkov effect is allowed, and thus $k \in [0,k_\text{th}]$. %
The process is permitted when $k_\text{th} > 0$, and the limit case $k_\text{th} = 0$ is given setting an equal sign in Eq. \eqref{eq:limit_k_0}.\par
%
%
When mass is considered in the previous analysis, one has to take into account that $\Theta_c(p) < 0$ always. %
The discussion when $\mathbb{A} < 0$ and $\mathbb{B} < 0$ holds: if $f(0) > 0$ there is only one solution to the thresholds condition and it is the maximum momentum that particle $b$ can have. %
In the case where $\mathbb{A} > 0$ and $\mathbb{B} > 0$, when $f(0) < 0$ it is possible to have two solutions, so that $k \in [k_\text{min}, k_\text{max}]$ with $k_\text{max} < p$, or no solutions.\par
%
%
\begin{table}[t]
    \centering
    \begin{tabular}{| >{\centering\arraybackslash}m{0.5cm} | >{\centering\arraybackslash}m{0.5cm} | >{\centering\arraybackslash}m{6.1cm}| >{\centering\arraybackslash}m{2.5cm} | >{\centering\arraybackslash}m{2.5cm} |}\hline
        $\alpha$ & $\beta$ & $k_\text{th}$ & $\mathbb{A},\mathbb{B} > 0$ & $\mathbb{A},\mathbb{B} < 0$ \\ \hline \vspace{7.5pt}
        $0$ & $0$ & no threshold & $\displaystyle \mathbb{B} - \mathbb{A} > 0$ & $\displaystyle \mathbb{B} - \mathbb{A} > 0$ \\[7.5pt]
        $0$ & $\mathbb{R}^+$ & $\displaystyle \left( \frac{\mathbb{A}}{\mathbb{B}} \right)^{1/\beta}$ & $\displaystyle \mathbb{B}p^\beta - \mathbb{A} > 0$ & $\displaystyle \mathbb{A} < - \frac{m^2}{2p^2}$ \\[7.5pt]
        $1$ & $0$ & $\displaystyle 2p - \frac{\mathbb{B}}{\mathbb{A}}$ & $\displaystyle \mathbb{B} - \mathbb{A}p > 0$ & $\displaystyle \mathbb{B} - 2\mathbb{A}p < \frac{m^2}{2p^2}$ \\[7.5pt]
        $1$ & $1$ & $\displaystyle \frac{2\mathbb{A}p}{\mathbb{A} + \mathbb{B}}$ & $\displaystyle \mathbb{B}p - \mathbb{A}p > 0$ & $\displaystyle \mathbb{A} < - \frac{m^2}{4p^3}$ \\[7.5pt]
        $1$ & $2$ & $\displaystyle \frac{1}{2\mathbb{B}} \left( -\mathbb{A} \pm \sqrt{\mathbb{A}^2 + 8\mathbb{A}\mathbb{B}} \right)$ & $\displaystyle \mathbb{B}p^2 - \mathbb{A}p > 0$ & $\displaystyle \mathbb{A} < - \frac{m^2}{4p^3}$ \\[7.5pt]
        $2$ & $0$ & $\displaystyle \frac{1}{2\mathbb{A}} \left( 3\mathbb{A}p \pm \sqrt{4\mathbb{A}\mathbb{B} - 3\mathbb{A}^2p^2} \right)$ & $\displaystyle \mathbb{B} - \mathbb{A}p^2 > 0$ & $\displaystyle \mathbb{B} - 3\mathbb{A}p^2 > \frac{m^2}{2p^2}$ \\[7.5pt]
        $2$ & $1$ & $\displaystyle \frac{1}{2\mathbb{A}} \left( 3\mathbb{A}p + \mathbb{B} \pm \sqrt{\mathbb{B}^2 + 6\mathbb{A}\mathbb{B}p - 3\mathbb{A}^2p^2} \right)$ & $\displaystyle \mathbb{B}p - \mathbb{A}p^2 > 0$ & $\displaystyle \mathbb{A} < - \frac{m^2}{6p^4}$ \\[7.5pt]
        $2$ & $2$ & $\displaystyle \frac{1}{2(\mathbb{A} - \mathbb{B})} \left( 3\mathbb{A}p \pm \sqrt{12\mathbb{A}\mathbb{B} - 3\mathbb{A}^2p^2} \right)$ & $\displaystyle \mathbb{B}p^2 - \mathbb{A}p^2 > 0$ & $\displaystyle \mathbb{A} < - \frac{m^2}{6p^4}$ \\[7.5pt]
        \hline
    \end{tabular}
    \caption{In this table we show the threshold momentum $k_\text{th}$ for the particle $b$ that satisfies the condition $f(k) = 0$ together with the conditions that $\mathbb{A}$ and $\mathbb{B}$ should satisfy for the process to be allowed. %
    The $\pm$ signs in the quadratic solutions depend on the signs of $\mathbb{A}$ and $\mathbb{B}$, and these are chosen so that $k_\text{th}$ is positive for those LIVPs that fulfill either the condition \protect\eqref{eq:limit_k_0} or \protect\eqref{eq:limit_k_p}. %
    We emphasize again that $k_\text{th}$ establishes the minimum momentum of the particle $b$ when $\mathbb{A}, \mathbb{B} > 0$, and a maximum momentum when $\mathbb{A}, \mathbb{B} < 0$.}
    \label{tab:threshold}
\end{table}
%
%
The threshold condition $\Theta_c = 0$ does not have a general solution $k_\text{th}$ for arbitrary $\alpha$ and $\beta$ since it is a polynomial Eq. of degree $\max\{ \alpha + 1, \beta +1 \}$. %
Nevertheless, particular solutions for small values of $\alpha$ and $\beta$ can be derived, and further simplified in the massless limit $m \to 0$. 
For example, consider the case where $\alpha = 1$ and $\beta = 0$. One easily checks from Eq. \eqref{eq:f(k)} that the threshold momentum for $m = 0$ is given by
\begin{equation}\label{eq:k_th_1_0}
    k_\text{th}(\alpha = 1, \beta = 0) =
    2p - \frac{\mathbb{B}}{\mathbb{A}}.
\end{equation}
Let us start discussing the case $\mathbb{A}, \mathbb{B} > 0$, where the threshold momentum is the minimum value of $k$ for the process to be allowed. %
Then, $k_\text{min} > p$ if $\mathbb{B} < \mathbb{A} p$ and the process is forbidden, as can be seen from Eq. \eqref{eq:limit_k_p}. %
Otherwise, the process is allowed with minimum momentum given by Eq. \eqref{eq:k_th_1_0}; in particular, the process is allowed for all $k$ when $\mathbb{B} > 2\mathbb{A}p$; see Eq. \eqref{eq:limit_k_0}. %
The analysis is reversed when considering $\mathbb{A}, \mathbb{B} < 0$. %
Now Eq. \eqref{eq:k_th_1_0} establishes the maximum momentum up to where the process is allowed. %
As long as $\mathbb{B} < 2\mathbb{A}p$ the process is permitted, and for $\mathbb{B} < \mathbb{A}p$ it is possible for all values of $k$; for $\mathbb{B} > 2\mathbb{A}p$ Cherenkov emission is not possible as $k_\text{max} < 0$. 
%
%
A similar analysis can be carried out solving Eq. \eqref{eq:threshold} for different values of $\alpha$ and $\beta$, and study whether the process is allowed or not attending to the signs of $\mathbb{A}$ and $\mathbb{B}$ and to Eqs. \eqref{eq:limit_k_0} and \eqref{eq:limit_k_p}. %
Table \ref{tab:threshold} illustrates the solutions to the threshold condition in the massless limit for the values $\alpha, \beta \in \{0,1,2\}$.\par%

\section{Dynamics of generalised Cherekov radiation}\label{sec_III}
%
%
In this Sec. we focus on the calculation of the rate of energy loss $\text{d}E/\text{d}t$ of the particle $a$ in the GCR process $a \to a + b$. %
We present the general procedure for the tree-level diagram shown in Fig. \ref{fig:generalized_cherenkov} for an arbitrary interaction vertex. Then we study the particular cases where the emitted particle is either a photon $\gamma$ (electromagnetic CR) or a graviton $h$ (gravitational CR).
%
%
\subsection{Decay rate and energy loss}\label{sec_III_a}
%
%
It is well known \cite{peskin1995introduction} that the differential decay rate of a two body process in an arbitrary Ref. frame in a LI scenario is given by
\begin{equation}\label{eq:decay_rate}
    \text{d}\Gamma =
    \frac{1}{8\pi^2 E_{\boldsymbol{p}}} \frac{\text{d}^3 \boldsymbol{p}^\prime}{2E_{\boldsymbol{p}^\prime}} \frac{\text{d}^3 \boldsymbol{k}}{2E_{\boldsymbol{k}}} \delta^{(4)} (p - p^\prime - k) \overline{|\mathcal{M} (p \to p^\prime, k)|^2}.
\end{equation}
Here $\boldsymbol{p}$ is the 3-momentum of the initial particle, $\boldsymbol{p}^\prime$ and $\boldsymbol{k}$ are the 3-momentum of the final particles, and $\mathcal{M} (p \to p^\prime, k)$ is the Lorentz invariant matrix element which depends on the interaction vertex considered. %
The Dirac delta function imposes the conservation of the energy and the 3-momentum, and the 3-momentum differentials come from the Lorentz invariant phase space (LIPS).\par
%
%
It is important to understand whether Eq. \eqref{eq:decay_rate} is allowed to be used in a LI breaking scenario, as is the case with our MDRs. %
When LI is not broken, the canonical commutation relations between creation and annihilation operators of the field are $[a_{\boldsymbol{p}}, a_{\boldsymbol{p}^\prime}^\dag] = (2\pi)^3 \delta^{(3)}(\boldsymbol{p} - \boldsymbol{p}^\prime)$ when
the usual factor $(2E_{\boldsymbol{p}})^{-1/2}$ in the momentum integral of the field operators is included.
This factor on the field operators is reflected in the structure of the LIPS in Eq. \eqref{eq:decay_rate}. %
This is no longer true when LI does not hold. %
Indeed, the wave function of the particles involved in the process will have a different normalization condition, including corrections due to the LI breaking term in the MDR. %
Hence, to ensure the canonical normalization condition, field operators will no longer be normalized by a factor $(2E_{\boldsymbol{p}})^{-1/2}$ and the LIPS in Eq. \eqref{eq:decay_rate} should receive additional corrections. %
Nevertheless, as we shall see during this Sec., the matrix elements $\mathcal{M}$ that we will consider violate LI, and so no corrections must be taken into account coming from the LIPS as long as we compute the decay rate in the preferred frame. %
Eq. \eqref{eq:decay_rate} is then valid for the purposes of this work \cite{Jacobson_2006}.\par
%
%
Let us now compute the integrals in Eq. \eqref{eq:decay_rate}. First of all, if we use the conservation of the 3-momentum we can immediately integrate over $\boldsymbol{p}^\prime$ and obtain
\begin{equation}
    \text{d}\Gamma =
    \frac{1}{8\pi^2 E_{\boldsymbol{p}}} \frac{1}{2E_{\boldsymbol{p} - \boldsymbol{k}}} \frac{\text{d}^3 \boldsymbol{k}}{2E_{\boldsymbol{k}}} \delta (E_{\boldsymbol{p}} - E_{\boldsymbol{p} - \boldsymbol{k}} - E_{\boldsymbol{k}}) \overline{|\mathcal{M} (p \to p^\prime, k)|^2},
\end{equation}
where $\mathcal{M}$ is assumed to be evaluated at $p^\prime = (E_{\boldsymbol{p} - \boldsymbol{k}}, \boldsymbol{p} - \boldsymbol{k})$. %
To integrate over $\boldsymbol{k}$ we can use spherical coordinates and fix the 3-momentum $\boldsymbol{p}$ along the $z$ axis. %
Note that this is possible since the MDR is invariant under rotations (and thus is the decay rate). %
As will be seen, the matrix element $\mathcal{M} (p \to p^\prime, k)$ for the processes we consider does not depend on the azimuthal angle, so
\begin{equation}
    \Gamma =
    \frac{1}{16\pi E_{\boldsymbol{p}}} \int \text{d}k\, k^2 \int_{-1}^1 \text{d}\cos{\theta}\, \frac{1}{E_{\boldsymbol{p} - \boldsymbol{k}} E_{\boldsymbol{k}}}  \delta (E_{\boldsymbol{p}} - E_{\boldsymbol{p} - \boldsymbol{k}} - E_{\boldsymbol{k}}) \overline{|\mathcal{M} (p \to p^\prime, k)|^2}.
\end{equation}
On the other hand, elemental properties of the Dirac delta function allow us to write
\begin{equation}\label{eq:propertie_delta}
    \delta (E_{\boldsymbol{p}} - E_{\boldsymbol{p} - \boldsymbol{k}} - E_{\boldsymbol{k}}) =
    2 E_{\boldsymbol{p} - \boldsymbol{k}} \delta (E_{\boldsymbol{p} - \boldsymbol{k}}^2 - [E_{\boldsymbol{p}} - E_{\boldsymbol{k}}]^2).
\end{equation}
When deriving the threshold condition in Sec. \ref{sec_II}, it can be shown from Eq. \eqref{eq:cos_2} that up to order $\mathcal{O}(\Theta_c)$
\begin{equation}
    E_{\boldsymbol{p} - \boldsymbol{k}}^2 - [E_{\boldsymbol{p}} - E_{\boldsymbol{k}}]^2 =
    2pk \left( \cos{\theta_c} - 1 + \Theta_c \right),
\end{equation}
which, substituting in Eq. \eqref{eq:propertie_delta}, gives
\begin{equation}
    \delta (E_{\boldsymbol{p}} - E_{\boldsymbol{p} - \boldsymbol{k}} - E_{\boldsymbol{k}}) =
    \frac{E_{\boldsymbol{p} - \boldsymbol{k}}}{pk} \delta (\cos{\theta_c} - 1 + \Theta_c).
\end{equation}
Finally, integrating over $\cos{\theta_c}$ one gets
\begin{equation}\label{eq:decay_rate_general}
    \Gamma =
    \frac{1}{16\pi p^2} \int \text{d}k\, H(\Theta_c) \overline{|\mathcal{M} (p \to p^\prime, k)|^2}.
\end{equation}
Note that we have approximated $E_{\boldsymbol{p}} = p + \mathcal{O}(\mathbb{A})$ in the denominator for the same reason why we did not consider higher order terms arising from the LIPS. %
The integration limits for $k$ are determined by the threshold condition \eqref{eq:threshold}, where $k = 0$ and $k = p$ the lowest and highest integration limits possible, respectively. %
$H(\Theta_c)$ is the Heaviside function and establishes whether the process is kinematically allowed or forbidden.\par
%
%
Eq. \eqref{eq:decay_rate_general} is valid for any process $a \to a + b$ with the MDR proposed with a matrix element $\mathcal{M}$ that does not depend on the azimuthal angle. %
However, the emission rate of the $b$ particles does not allow us to impose any constraint on the LIVPs, unlike the energy loss rate of the $a$ particle. %
To consider the energy carried off by the particle $b$, we must insert the energy $E_{\boldsymbol{k}} = k + \mathcal{O}(\mathbb{B})$ of this particle in the integral of Eq. \eqref{eq:decay_rate_general}, obtaining the following energy loss rate of the particle $a$:
\begin{align}\label{eq:energy_loss_rate_general}
    \frac{\text{d} E}{\text{d} t} =
    \frac{1}{16\pi p^2} \int\text{d}k\, k H(\Theta_c) \overline{|\mathcal{M} (p \to p^\prime, k)|^2}.
\end{align}
To evaluate the matrix element $\mathcal{M}$ we will consider, for simplicity, the case where the particle $a$ is a complex scalar field, even though the constraints obtained in Sec. \ref{sec_IV} for the LIVPs are applied to fermionic particles. %
It is expected that spin corrections are of order $\mathcal{O}(1)$ and, therefore, are not significant enough to modify the order of magnitude of the constraints \cite{Moore_2001}.\par
%
%
\subsection{Electromagnetic Cherenkov radiation}\label{sec_III_b}
%
%
Let us first consider the process described in Fig. \ref{fig:generalized_cherenkov} where the particle $b$ is a photon $\gamma$. %
Feynman rules for a charged complex scalar field applied to this process give the following matrix element \cite{schwartz2013quantum}:
\begin{equation}
    i\mathcal{M} = -ie_a(p^\mu + p^{\prime \mu}) \epsilon_\mu(k).
\end{equation}
Here $\epsilon_\mu$ is the polarization vector of the photon, and $e_a$ the electric charge of the particle $a$. %
The analysis we will carry out does not distinguish between polarizations; hence, we must sum over the two transverse polarization states of the photon when computing the squared matrix element. %
Using the conservation of 3-momentum we get
\begin{equation}
    \overline{|\mathcal{M} (p \to p^\prime, k)|^2} =
    \sum_\epsilon e_a^2 \left| (2p^\mu - k^\mu) \epsilon_\mu(k) \right|^2.
\end{equation}
The physical polarizations are perpendicular to the 4-momentum of the photon and therefore $k^\mu \epsilon_\mu(k) = 0$. %
The product $p^\mu \epsilon_\mu(k)$ is easily calculated taking into account that the two polarizations are perpendicular to each other and the angle between $\boldsymbol{p}$ and $\boldsymbol{k}$ is $\theta_c$. %
This yields
\begin{equation}\label{Mem}
    \overline{|\mathcal{M} (p \to p^\prime, k)|^2} =
    4e_a^2 p^2 \sin^2{\theta_c} =
    8e_a^2 p^2 \Theta_c + \mathcal{O}(\Theta_c^2),
\end{equation}
where we have used Eq. \eqref{eq:cos_final}, considered the LI violating terms up to first order, and again denoted $p = |\boldsymbol{p}|$. %
Substituting this matrix element in Eq. \eqref{eq:energy_loss_rate_general}, the electromagnetic energy loss rate is then given by
\begin{equation}\label{eq:energy_loss_rate_EM}
    \frac{\text{d} E}{\text{d} t} =
    \frac{e_a^2}{2\pi} \int\text{d}k\, k \Theta_c(k) H(\Theta_c).
\end{equation}
The integral has a simple solution:
\begin{align}
    \frac{\text{d}E}{\text{d}t} =
    \frac{e_a^2}{2\pi} \Bigg[ &\mathbb{A}p^{\alpha + 2} \left( 1 - \frac{k}{p} \right)^2 \left\{ \frac{1}{2} - \frac{1}{\alpha + 3} \left( 1 - \frac{k}{p} \right)^{\alpha + 1} \right\} \notag \\ +& \mathbb{B}_\gamma k^{\beta + 2} \left\{ \frac{1}{\beta + 2} - \frac{1}{\beta + 3} \frac{k}{p} \right\} - \frac{m^2 k^2}{4p^2} \Bigg]_{k_\text{min}}^{k_\text{max}},
\end{align}
where the integration limits $k_\text{min}$ and $k_\text{max}$ are obtained by solving the threshold condition \eqref{eq:threshold}. %
Also, note that if $\Theta_c(k) < 0$ for all $k \in [0,p]$ the Heaviside function in Eq. \eqref{eq:energy_loss_rate_EM} imposes the energy loss rate to be equal to zero, as Cherenkov emission is kinematically forbidden.\par
%
%
The key feature of this emission rate is that it is of first order in $\Theta_c$, and hence in the LI breaking parameters, $\mathbb{A}$ and $\mathbb{B}$, and in $m^2/p^2$. %
We have checked that the results obtained here coincide with those in \cite{Moore_2001} when considering $\mathbb{A} = 0$ and $\beta = 0$. %
Similar MDRs have been used to study vacuum electromagnetic CR \cite{Jacobson_2003, Jacobson_2006} and it was shown that for $\alpha = \beta \neq 0$ a significant fraction of the energy of the particle $a$ is emitted almost immediately when $a$ has an energy above the threshold of the process. %
In Sec. \ref{sec_IV.C} we will also check with a simple estimation that when $\alpha \neq \beta$ and neither are equal to $0$, we obtain the same result. %
In particular, when $\alpha = \beta = 0$, the energy loss rate is suppressed by the difference of the maximum attainable velocities of the particles, $\text{d}E/\text{d}t \sim (c_a^2 - c_\gamma^2) \alpha_\text{em} E^2$, where $\alpha_\text{em}$ is the fine structure constant \cite{Coleman_1999}. %
Thus, a kinematical analysis is enough to establish stringent constraints to LIVPs and it is not necessary to consider propagation effects, i.e., compute the energy loss rate. %
In this scenario the study of the threshold condition in Sec. \ref{sec:II_b} should be carried out obtaining the threshold momentum of the particle $a$, namely $p_\text{th}$. %
Constraints can then be imposed using high energy astrophysical observations and considering the energy of the particle $a$ detected as the threshold energy of the process. %
This has been done for $\alpha = \beta \in \mathbb{Z}^+$ \cite{Coleman_1997, Coleman_1999, Jacobson_2003, Jacobson_2006} but a generalization to $\alpha \neq \beta$ solving the threshold condition \eqref{eq:threshold} for the momentum of the particle $a$ could be performed for small values of the exponents $\alpha$ and $\beta$.
%
%
\subsection{Gravitational Cherenkov radiation}\label{sec_III_c}
%
%
We now consider the case where the particle $b$ in Fig. \ref{fig:generalized_cherenkov} is a graviton $h$. %
The Feynman rules for the process $a \to a + h$, with $a$ a complex scalar field can be found in Ref. \cite{Han_1999}:
\begin{eqnarray}
    i\mathcal{M} &=&
    -i\sqrt{4\pi G_\text{N}} (m^2 \eta_{\mu\nu} + C_{\mu\nu \rho\sigma} p^\rho p^{\prime\sigma}) \epsilon^{\mu\nu}, \label{eq:feynman_rule_M}
    \\
     C_{\mu\nu \rho\sigma} &=&
    \eta_{\mu\rho} \eta_{\nu\sigma} + \eta_{\mu\sigma} \eta_{\nu\rho} - \eta_{\mu\nu} \eta_{\rho\sigma}. \label{eq:feynman_rule_C}
\end{eqnarray}
Here $G_\text{N}$ is the gravitational constant, and $\epsilon^{\mu\nu}$ is the polarization tensor for the graviton field $h_{\mu\nu}$. %
This tensor is traceless ($\epsilon_{\mu\nu} \eta^{\mu\nu} = 0$) and transverse ($\epsilon_{\mu\nu} k^\mu = 0$), and can be constructed in terms of the polarization vectors of massive vector bosons \cite{Han_1999}. %
As in the electromagnetic case, we do not distinguish between the polarizations of the graviton and thus we sum over the two physical polarization states. Using Eq. \eqref{eq:feynman_rule_C} and substituting in the matrix element \eqref{eq:feynman_rule_M} one finds
\begin{equation}\label{eq:matrix_element_grav}
    \overline{|\mathcal{M} (p \to p^\prime, k)|^2} =
    16\pi G_\text{N} \sum_\epsilon |p_\mu p_\nu \epsilon^{\mu\nu}|^2 =
    16\pi G_\text{N} p^4 \sin^4{\theta_c}.
\end{equation}
Comparing this expression with that corresponding to the electromagnetic CR process, Eq. \eqref{Mem}, it can be noted that for the gravitational case instead of the electric charge $e_a^2$ we have the gravitational constant times the 3-momentum squared $G_\text{N} p^2$, and the emission is reduced for small opening angles, having $\sin^4{\theta_c}$ rather than $\sin^2{\theta_c}$. %
Substituting the matrix element \eqref{eq:matrix_element_grav} in the energy loss rate \eqref{eq:energy_loss_rate_general} and considering the lowest order in $\Theta_c$ in Eq. \eqref{eq:cos_final} we obtain
\begin{equation}\label{eq:energy_loss_rate_grav}
    \frac{\text{d} E}{\text{d} t} =
    4 G_\text{N}p^2 \int \text{d}k\, k\, \Theta_c^2(k) H(\Theta_c).
\end{equation}
%
%
We first notice that the energy loss rate \eqref{eq:energy_loss_rate_grav} of the vacuum gravitational CR, due to the tensor nature of the gravitational field $h$, is of order $\mathcal{O}(\Theta_c^2)$, whereas the energy loss rate \eqref{eq:energy_loss_rate_EM} of the electromagnetic CR is of order $\mathcal{O}(\Theta_c)$ because of the vector nature of the EM field. %
On the other hand, the electromagnetic coupling constant is much stronger than the gravitational coupling for energies much lower than the Planck scale: the factor $e_a^2/(2\pi)$ in Eq. \eqref{eq:energy_loss_rate_EM} is typically of order $\mathcal{O}(10^{-2})$, while in contrast the factor $G_\text{N} p^2$ in \eqref{eq:energy_loss_rate_grav} is of order $\mathcal{O}(10^{-14})$ for the highest energy particles observed in the Universe. %
Therefore, as expected, the energy loss rate due to gravitational CR is much smaller than that caused by the emission of electromagnetic CR. %
So, for gravitational CR the decay rate of a particle of astrophysical origin can be comparable to its travel time and, therefore, it must be taken into account in order to constrain the LIVPs. %
Unlike for the electromagnetic CR, a kinematic analysis is not enough to determine an upper bound for $\mathbb{A}$ and $\mathbb{B}$.\par
%
%
We shall also remark that it is possible to have both electromagnetic and gravitational CR emission at the same time. %
In that scenario, the electromagnetic process is much more efficient, as we have discussed, and it is the dominant source of energy loss. %
Hence, constraints coming from gravitational CR can be considered as long as the LIVP for the photon $\mathbb{B}_\gamma$ is such that the electromagnetic process is always forbidden.\par
%
%
The indefinite integral in Eq. \eqref{eq:energy_loss_rate_grav} can be computed in the general case $m \neq 0$ and $\alpha, \beta \geq 0$ in terms of hypergeometric functions, although it is not particularly illuminating. %
In addition, we also need to calculate the threshold momentum $k_\text{th}$ for the integration limits, which cannot be done analytically for arbitrary values of $\alpha$ and $\beta$. %
Nevertheless, the energy loss rate of the $a$ particle can be obtained for particular values of the parameters. For example, taking $\mathbb{A} = 0$, one gets in the massless limit
\begin{equation}
    \frac{\text{d}E}{\text{d}t} =
    \frac{2G_\text{N} \mathbb{B}^2 p^{2(\beta + 2)}}{(\beta + 2)(\beta +1) (2\beta + 3)},
\end{equation}
as it was obtained in Ref. \cite{Kiyota_2015}%
\footnote{This result corresponds to Eq. (13) of \cite{Kiyota_2015} taking $\delta = 0$ and identifying our exponent $\beta$ with $\alpha - 2$ (do not confuse this exponent with that of the particle $a$) and our parameter $\mathbb{B}$ with $-A/2$.} for $k_\text{min} = 0$ and $k_\text{max} = p$. %
This result reduces to that presented in Ref. \cite{Moore_2001} if we additionally impose $\beta = 0$.

\section{Constraints on the Lorentz invariance violating parameters}\label{sec_IV}
%
%
We are now interested in obtaining estimations for the LIVPs $\mathbb{A}$ and $\mathbb{B}$ appearing in $n_a$ and $n_b$ using the energy loss rate of vacuum gravitational CR \eqref{eq:energy_loss_rate_grav}. %
We shall use high energy astrophysical observations for this purpose, arguing that UHECRs offer the most stringent bounds. %
We will consider different values of $\alpha$ and $\beta$ to discuss the corresponding phenomena and illustrate it in Fig. \ref{fig:constraints_CR} for $\alpha = 2$, $\beta = 0$ and $\alpha = 1$, $\beta = 2$.\par
%
%
\subsection{Constraints from the energy loss rate}
%
%
In Sec. \ref{sec_III} we discussed that the decay rate of a particle of astrophysical origin due to gravitational CR is expected to be of the same order as the time of propagation, which implies that the energy loss rate has to be taken into account in order to estimate the LIVPs of both the particle $a$ and the theory of gravity. %
Computing the integral in Eq. \eqref{eq:energy_loss_rate_grav} and solving the differential Eq. would give the maximum time travel $t$ possible for a given momentum $p$ in terms of $\mathbb{A}$, $\mathbb{B}$ \cite{Moore_2001, Kimura_2012}. %
Thus, constraints on $\mathbb{A}$ and $\mathbb{B}$ can be imposed using high energy astrophysical observations if the distance traveled $ct$ and the momentum $p$ are known.\par
%
%
However, we have argued that it is not possible to obtain the maximum time travel $t$ for arbitrary values of $\alpha$ and $\beta$ in terms of $\mathbb{A}$ and $\mathbb{B}$, as $k_\text{th}$ cannot be computed analytically for arbitrary values of $\alpha$ and $\beta$. %
In this scenario, the condition at which damping from gravitational CR becomes relevant for a particle with energy $p$ travelling for a time $t$ may be estimated as $\text{d}E/\text{d}t \gg p/t$ \cite{Kiyota_2015}. %
Then, as the $a$ particles arrive to the Earth, we establish that values of $\mathbb{A}$ and $\mathbb{B}$ that satisfy $\text{d}E/\text{d}t \gg p/t$ are excluded by gravitational CR, and those that satisfy $\text{d}E/\text{d}t \ll p/t$ are allowed by observation.\par
%
To illustrate how constraints can be estimated, let us consider the simple case where $\alpha = \beta = 0$. %
Attending to Eqs. \eqref{eq:limit_k_0} and \eqref{eq:limit_k_p} we see that the process is allowed when $\mathbb{B} \geq \mathbb{A}$. %
Note that this condition is equivalent to the particle $a$ having a greater phase velocity than the particle $b$, $c_a > c_b$. %
Integration of Eq. \eqref{eq:energy_loss_rate_grav} in the massless limit is straightforward and yields
\begin{equation}\label{eq:alpha_beta_cero_m}
    \frac{\text{d}E}{\text{d}t} =
    \frac{G_\text{N}}{3} p^4 (\mathbb{A} - \mathbb{B})^2,
\end{equation}
so damping from CR is not significant if
\begin{equation}\label{eq:constraint_A_B_0_0}
    0 \leq \mathbb{B} - \mathbb{A} \ll \sqrt{\frac{3}{G_N t p^3}}.
\end{equation}
%
%
Here we see that constraints are more stringent for high energy particles that have traveled a long distance $ct$; note, however, that the energy of the particle plays a more relevant role since the dependence is of the form $(t p^3)^{-1/2}$. %
Another interesting feature is that the constraint is only imposed for the difference between the LIVPs, so one of them remains unfixed: as long as this difference is small enough, the values of $\mathbb{A}$ and $\mathbb{B}$ can be arbitrarily large. %
One should be careful with this last statement as our results have been derived assuming that $|\mathbb{A} p^\alpha|$ and $|\mathbb{B} k^\beta|$ are very small compared to unity. %
We have checked that both of these features, the dependence on $p$ and $t$ and the impossibility of fixing both $\mathbb{A}$ and $\mathbb{B}$ simultaneously, are present for different values of $\alpha$ and $\beta$.\par
%
%
It is also worth mentioning the case where $\mathbb{A} = 0$ and $\beta$ is a positive real number. %
Performing a similar calculation in the massless limit $m = 0$ shows that damping from gravitational CR is not relevant if
\begin{equation}
    0 < \mathbb{B}^{(\beta)} p^\beta \ll
    \frac{\sqrt{(\beta + 1)(\beta + 2)(2\beta + 3)}}{\sqrt{2G_\text{N} tp^3}},
\end{equation}
which corresponds to the result derived in Ref. \cite{Kiyota_2015}. %
The constraint on $\mathbb{B}^{(\beta)} p^\beta$ exhibits the same dependence on the energy and the time travel as Eq. \eqref{eq:constraint_A_B_0_0}, and we see that the value of $\mathbb{B}^{(\beta)}$ decreases for higher values of $\beta$. %
This has been used to derive much more stringent constraints in the gravitational sector using the absence of vacuum CR for $\beta \gtrsim 0$ rather than using direct detection of GWs, in particular the events GW150914 and GW151226 \cite{Yunes_2016}. %
The same behavior is observed when the dispersion relation of $a$ particles is modified, that is, when $\mathbb{A} \neq 0$.
\subsection{Observational constraints in the gravitational sector}
UHECRs \cite{Abreu_2022} are particles with energies above $1\, \text{EeV}$ ($10^9\, \text{GeV}$), whose origin is most likely extragalactic and are mainly composed of protons and heavy nuclei \cite{Globus_2015}. %
At such high energies these particles interact with the intergalactic photon background through the Greissen-Zatsepin-Kuz'min effect: protons may lose energy due to photo-pion production, and heavy nuclei through photodissociation interaction. %
In both cases the mean free path is of order $200-300\, \text{Mpc}$ for energies around $100\, \text{EeV}$ \cite{Addazi_2022}. %
UHECRs offer the best scenario to test LI using CR; in fact, these are the most energetic particles observed in the Universe and are sensitive to the decay time of gravitational CR thanks to their long propagation distances.\par
%
%
In order to obtain realistic bounds \cite{Kosteleck_2015}, let us consider that the UHECR is a proton of energy $p \sim 10^{11}\, \text{GeV}$. %
The source of these particles is still unknown \cite{Abreu_2022} but it is expected that they are produced in active galactic nuclei; the nearest is found at a few Mpc, so we may take $ct \sim 10\, \text{Mpc}$. %
Here we do not take into account the spin of the proton and neither its inner structure. %
Regarding the former, we already have mentioned that spin corrections are not expected to affect the order of magnitude of the constraints on $\mathbb{A}_\text{proton} \equiv \mathbb{A}$ and $\mathbb{B}_\text{graviton} \equiv \mathbb{B}$. %
Attending to the latter, a detailed analysis of the gravitational CR emitted by a proton would require us to consider its partonic structure. %
For soft emitted gravitons with $k \to 0$, this inner structure might be neglected and the pointlike approximation is valid \cite{Jacobson_2006}; for hard emitted gravitons with $k \lesssim p$ it becomes important and the emission rate of its constituents should be computed. %
This would substantially increase the number of LIVPs considered as $\mathbb{A}_\text{quark} \neq \mathbb{A}_\text{gluon}$ in general, making the analysis more involved. %
This study is out of the scope of this work, but it is worth mentioning that this has been studied with more simple MDRs which consider $\alpha = \beta = 0$ \cite{Gagnon_2004,Elliott_2005}. %
In short, we will assume that the proton has its own effective MDR, whose structure could be understood in terms of the composition of the MDRs of its constituents.\par
\begin{figure*}[t]
    \centering
    \includegraphics[width=\linewidth]{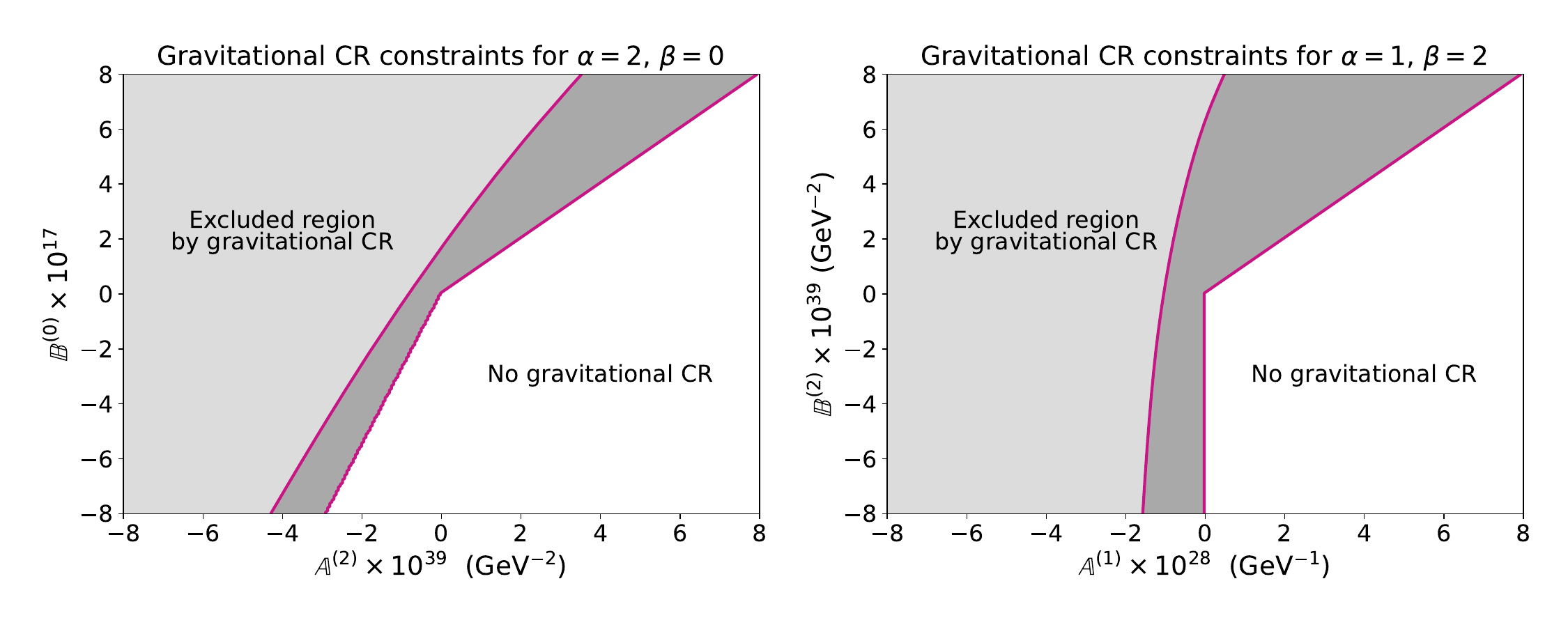}
    \vspace{-0.75cm}
    \caption{Constraints on the model for $\alpha = 2$, $\beta = 0$ (left panel) and $\alpha = 1$, $\beta = 2$ (right panel) using an UHECR with $p \sim 10^{11}\, \text{GeV}$ and $ct \sim 10\, \text{Mpc}$. The dark-gray region in both figures are the values of $\mathbb{A}_\text{proton} \equiv \mathbb{A}$ and $\mathbb{B}_\text{graviton} \equiv \mathbb{B}$ that satisfy the condition $\text{d}E /\text{d}t \ll p /t$. There, the damping of the UHECR due to gravitational CR is not significant, and it is possible for the particle to reach the Earth while emitting gravitons. The light-gray region shows that Cherenkov emission is kinematically allowed but does not satisfy $\text{d}E /\text{d}t \ll p /t$; hence, our model excludes those values of the LIVPs. The white region corresponds to $(\mathbb{A},\mathbb{B})$ where CR is not permitted, as $\Theta_c < 0$ for all possible values of $k$.}
    \label{fig:constraints_CR}
\end{figure*}

%
%
Computing the threshold momentum $k_\text{th}$ numerically for $m = m_\text{proton}$, performing the integral in Eq. \eqref{eq:energy_loss_rate_grav} and imposing the condition $\text{d}E/\text{d}t \ll p/t$ we have obtained the allowed $\mathbb{A}-\mathbb{B}$ parameter space for $\alpha = 2$, $\beta = 0$ and $\alpha = 1$, $\beta = 2$ shown in Fig. \ref{fig:constraints_CR}. %
Values of $\mathbb{A}$ and $\mathbb{B}$ that define the dark-grey region permit Cherenkov emission without significant damping, whereas in the light-grey region the process is kinematically allowed but forbidden by the observation because of the damping effect. %
In both cases we see that no constraints are imposed in the region with $\mathbb{A} > 0$ and $\mathbb{B} < 0$, and also when $\mathbb{A}, \mathbb{B} > 0$ and $\mathbb{B} p^\beta - \mathbb{A} p^\alpha < 0$. %
The latter is due to Eq. \eqref{eq:limit_k_p}, since the threshold momentum is $k_\text{min} > p$. %
When both $\mathbb{A}$ and $\mathbb{B}$ are negative the region where kinematics does not permit Cherenkov emission is different in both cases. %
Attending to Eq. \eqref{eq:limit_k_0}, the case where $\alpha = 2$ and $\beta = 0$ establishes that $\mathbb{B} - 3\mathbb{A}p^2 \geq 0$ so that the threshold momentum is $k_\text{max} > 0$; by contrast, when $\alpha = 1$ and $\beta = 2$ the maximum momentum exists when $\mathbb{A} \leq 0$.\par
%
%
%
%
The estimated excluded region appears in both cases for $|\mathbb{A} p^\alpha| \sim |\mathbb{B} p^\beta| \lesssim 10^{-17}$ when $p \sim 10^{11}\, \text{GeV}$ and $ct \sim 10\, \text{Mpc}$. %
Other positive integer values of $\alpha$ and $\beta$ exhibit constraints of the same order of magnitude for these energies and traveled distances. %
This allows us to check that LIVPs significantly modify the threshold condition of the process since $|\mathbb{A} p^\alpha| \gg (m/p)^2$, see Eq. \eqref{eq:deviation_threshold}. %
Hence, the approximations performed in Sec. \ref{sec_II} are consistent and the analyses in the massless limit are accurate.\par
%
%
As we have previously advanced, Fig. \ref{fig:constraints_CR} shows that it is possible to have arbitrarily large values of $\mathbb{A}$ and $\mathbb{B}$ where vacuum gravitational CR is allowed by observation as long as its difference is small enough. %
We have checked that the same happens for many other positive integer values of $\alpha$ and $\beta$. %
This fact can be understood in the following way. %
When computing the energy loss rate, Eq. \eqref{eq:energy_loss_rate_grav}, it is necessary to obtain the integration limits from the threshold condition $\Theta_c = 0$ for some given $\mathbb{A}$ and $\mathbb{B}$. %
Considering values of $\mathbb{A}$ and $\mathbb{B}$ where the process is allowed for some $k$, an increase in $\mathbb{A}$ and $\mathbb{B}$ will cause an increase in the value of $\Theta_c$ and thus in $\text{d}E/\text{d}t$. %
However, if the difference between $\mathbb{A}$ and $\mathbb{B}$ is small enough, the region of $k$ where the Cherenkov emission is possible will be narrower; this reduces the value of $\text{d}E/\text{d}t$ as the integration interval is smaller, and it can be such that the condition $\text{d}E/\text{d}t \ll p/t$ is still satisfied. %
Remember that $\Theta_c$ was obtained assuming $|\mathbb{A} p^\alpha| \sim |\mathbb{B} k^\beta| \ll 1$, and thus our model is not predictive for large values of $\mathbb{A}, \mathbb{B}$.\par
%
%
In this analysis we notice that when $\beta > 0$, given a small value of $\mathbb{A} < 0$ we obtain an upper bound for $\mathbb{B}$ but not a lower bound. %
This is not the case for positive values of $\mathbb{A}$ and $\mathbb{B}$, since fixing one of them constraints the other LIVP; the same happens for negative values of $\mathbb{A}$ and $\mathbb{B}$ when $\beta = 0$. %
Therefore, constraints derived for $\mathbb{A}$ from a different process may not fix $\mathbb{B}$ for $\beta > 0$. %
Note, however, that given a value of $\mathbb{B}$, determined by direct observation of GWs, it is always possible to obtain an upper bound for $\mathbb{A}$. %
Unfortunately, these measures of the parameter $\mathbb{B}$ are not stringent, so the bounds obtained for $\mathbb{A}$ are not competitive with those obtained from threshold constraints in the QED sector \cite{Jacobson_2003,Jacobson_2006}.
%
%
\subsection{Travel distance for particles emitting electromagnetic Cherenkov radiation}\label{sec_IV.C}
%
%
We have previously discussed that for positive integer values of $\alpha = \beta$ the energy loss rate of the particle $a$ due to vacuum electromagnetic CR is large enough so that the traveled distance for energies above the threshold $p_\text{th}$ is negligible compared to astrophysical distances \cite{Coleman_1997, Coleman_1999, Jacobson_2003, Jacobson_2006}. %
As our MDR does not assume equal values of the exponents of the particle $a$ and the photon ($\alpha$ and $\beta_\gamma$, respectively), it could be possible for $a$ to emit CR over large distances. %
Here we show that, even when $\alpha \neq \beta$, the vacuum electromagnetic CR is much more efficient than its gravitational counterpart, and that the particle $a$ loses all of its energy almost immediately above the threshold.\par
%
%
\begin{figure*}[t]
    \centering
    \includegraphics[width=\linewidth]{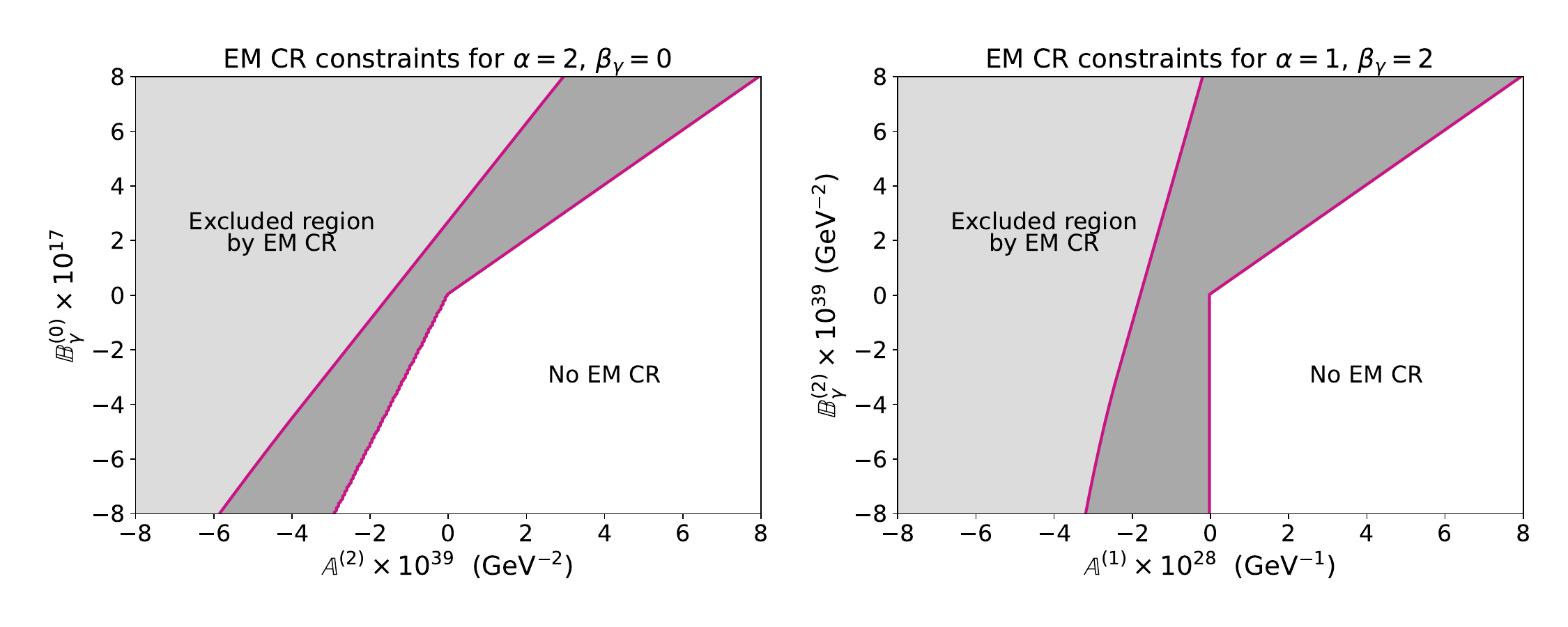}
    \vspace{-0.75cm}
    \caption{Constraints on the model for $\alpha = 2$, $\beta = 0$ (left panel) and $\alpha = 1$, $\beta_\gamma = 2$ (right panel) using an UHECR with $p \sim 10^{11}\, \text{GeV}$ and $ct \sim 0.1\, \text{fm}$. White, light-gray and dark-gray regions represent the same as in Fig. \ref{fig:constraints_CR} but for the existence or absence of vacuum electromagnetic CR.}
    \label{fig:constraints_EM_CR}
\end{figure*}
%
%
To illustrate this, let us consider an ultrahigh energy proton of energy $p \sim 10^{11}\, \text{GeV}$. %
As we do not solve the threshold Eq. \eqref{eq:threshold} to find the threshold momentum $p_\text{th}$ of the particle $a$, we estimate that the order of magnitude of $\mathbb{A}$ and $\mathbb{B}_\gamma$ given by the kinematical constraints is the same as the one obtained in the gravitational case; see Fig. \ref{fig:constraints_CR}. %
For the same values of the exponents $\alpha$ and $\beta_\gamma$, Fig. \ref{fig:constraints_EM_CR} shows the parameter space $\mathbb{A} - \mathbb{B}_\gamma$ assuming a traveled distance of $ct \sim 0.1\, \text{fm}$. %
We see that the region where the damping is not significant lies in distance scales $39$ orders of magnitude smaller compared to those assumed for the emission of vacuum gravitational CR. %
This means that for UHECRs whose origin is galactic or extragalactic, the emission of electromagnetic CR occurs almost immediately after being produced, losing energy until reaching the threshold momentum $p_\text{th}$: the dark-grey region occupies an insignificant amount of the parameter space, and thus the kinematical analysis is enough to obtain bounds on the LIVPs. %
The estimation was made for two specific pair of values $(\alpha, \beta)$ but we have checked explicitly that similar results are obtained for many other positive integer values.\par

\section{Conclusions}\label{sec_V}
%
%
Vacuum Cherenkov emission occurs as an emission process of a massless particle and it is possible when LI is broken \cite{Schreck_2018, Moore_2001, Mattingly_2005, Coleman_1997, Coleman_1999, Jacobson_2003, Jacobson_2006, Gagnon_2004, Elliott_2005, Kimura_2012, Kiyota_2015, Kosteleck_2015}. %
An efficient way to implement LIV is through MDRs, which are able to encapsulate QG phenomenology \cite{Addazi_2022, batista2023white}. %
In this work we have proposed a new class of LI violating MDRs inspired on the classical electromagnetic Cherenkov emission in an optical medium with refractive index $n$, different for each particle species and with a power law dependence on the energy. %
Kinematics of the GCR process have been studied and the momentum configurations for which the process is allowed have been derived. %
These conditions were implemented to obtain the energy loss rate due to GCR, paying particular attention to the electromagnetic and gravitational vacuum Cherenkov radiation. %
The latter results served to estimate the LIVPs of UHECRs.\par
%
%
We would like to remark on an important feature: the parameter space of the LI violating terms is not bounded {\color{Red} from above and below, i.e., it is not closed (note that from Fig. \ref{fig:constraints_CR}, $\mathbb{A}$ is bounded from below and $\mathbb{B}$ from above)}, which implies that gravitational vacuum Cherenkov radiation is permitted as long as the difference between them is small enough. %
Different processes, such as the decay of a graviton in two high energy particles, could be used to restrict the allowed parameter space, but the quantum nature of gravity has not been resolved yet and no observations of gravitons have been made. %
Note that this problem is not present in the QED sector. %
On the one hand, the characteristic time of the analog processes mentioned is much smaller than the traveled distance of astrophysical particles; hence, constraints can be imposed using only kinematics. %
On the other hand, electrons and photons of high energy have been detected, and thus the parameter space of the LIVPs of these particles can be restricted from the absence of the vacuum electromagnetic CR and photon decay processes, or the change in the threshold of the photon annihilation process \cite{Jacobson_2003,Jacobson_2006}.\par
%
%
{\color{Red}
The study carried out along this work is applicable for arbitrary values of the exponents $\alpha$ and $\beta$ until we are interested in deriving additional kinematic features (the characteristics of the solutions to the threshold equation) when both LIVPs $\mathbb{A}$ and $\mathbb{B}$ have equal signs. %
From this point on we focus on positive values of $\alpha$ and $\beta$, i.e., in violation of LI relevant in the ultraviolet, as expected from many approaches to quantum gravity \cite{Addazi_2022,batista2023white}. %
Nevertheless, a scenario in which the exponent in the MDR is negative has been studied in \cite{Freidel_2021} and Ref.s therein, arguing that quantum gravity phenomenology in the infrared might be possible. %
In such a case, the phenomenology of UHECR is not relevant for searching of possible signatures of such a scenario. Thus, infrared MDRs do not modify significantly the kinematics of the signatures studied in this work. %
On the other hand, vacuum gravitational CR associated to low energy processes could serve to impose constraints in the LIVP of the graviton, $\mathbb{B}^{(\beta)}$, when $\beta < 0$, as it is possible that the threshold condition admits solutions for arbitrary low momenta depending on the particular MDR under consideration. %
This interesting study is out of the scope of this work, but we look forward for future projects tackling this aspect.\par%
}
%
%
It should be noted that even though the present work is focused on the study of threshold effects and more specifically particle decays, MDRs have a much broader scope to restrict LIVPs. %
Constraints in the gravitational sector may be imposed using time of flight delay in multimessenger detections. %
At the present only one multimessenger detection mixing GWs and electromagnetic radiation has been reported, the GW signal GW170817 with the gamma-ray signal from a kilonova \cite{Abbott_2017}, but these restrictions using the time delay between GWs and photons have been applied only to impose limits on the Standard Model extension parameters in Ref. \cite{Kosteleck_2015}. %
Another possibility, which is not considered in the MDR proposed here, is the vacuum birefringence phenomenon caused by the difference in the propagation velocity between the states $h_+$ and $h_\times$ of GWs.\par
%
%
In conclusion, our work establishes a systematic method to study two body particle decays in a LI violating scenario and impose constraints through high energy astrophysical observations. %
Extensions to more particle interaction processes can be carried out following the steps exemplified, but these are not of much interest in the gravitational sector. %
We also want to emphasize the importance of using complementary observations, such as direct detection of GWs, in order to improve the bounds on the parameters introduced. %
This will be reinforced with future detections of UHECRs and multimessenger events, which may allow us to understand the elusive nature of QG.

\bigskip
\acknowledgments
    P.M.M. would like to thank Iarley P. Lobo for pointing out to her interesting Ref.s. %
    The research of P.M.M. is supported by the project PID2022-138263NB-I00 funded by MICIN/AEI/10.13039/ 501100011033  and by ERDF/EU. The work of J.A.R.C. is partially supported by the project PID2022-139841NB-I00 funded by MICIU/AEI/10.13039/501100011033 and by ERDF/EU, and by COST
(European Cooperation in Science and Technology) Actions CA22113 and CA23130. M.A. acknowledges financial support through IPARCOS Master grant.

\bibliographystyle{jhep}
\bibliography{bibliography}
\end{document}